\documentclass[preprint,floats,prl,showpacs]{revtex4-1}

\usepackage{amsmath}
\usepackage{amsfonts}

\usepackage{tikz}

\newcommand {\be}{\begin{equation}}
\newcommand {\ee} {\end{equation}}
\newcommand {\bea}{\begin{eqnarray}}
\newcommand {\eea} {\end{eqnarray}}

\begin{document}


\title{Overlap Structure and Free Energy Fluctuations in Short-Range Spin Glasses}

\author{C.M.~Newman$^{1,2,3}$ and D.L. Stein$^{1,2,4,5,6,7}$}
\affiliation{$^1$Courant Institute of Mathematical Sciences, New York University, New York, NY 10012, 
USA\\
$^2$NYU-ECNU Institute of Mathematical Sciences at NYU Shanghai, 3663 Zhongshan Road North, Shanghai, 
200062, China\\
$^3${\tt newman@cims.nyu.edu}\\
$^4$Department of Physics, New York University, New York, NY 10012, USA\\
$^5$NYU-ECNU Institute of Physics at NYU Shanghai, 3663 Zhongshan Road North, Shanghai, 200062, 
China\\
$^6$Santa Fe Institute, 1399 Hyde Park Rd., Santa Fe, NM 87501, USA\\
$^7${\tt daniel.stein@nyu.edu}}

\begin{abstract}
We investigate scenarios in which the low-temperature phase of short-range spin glasses comprises thermodynamic states which are
nontrivial mixtures of multiple incongruent pure state pairs. We construct a new kind of metastate supported on Gibbs states whose edge
overlap values with a reference state fall within a specified range. Using this metastate we show that, in any dimension, the variance of 
free energy difference fluctuations between pure states within a single mixed Gibbs state with multiple edge overlap values diverges 
linearly with the volume.  We discuss some implications of these results.

\end{abstract}


\maketitle


There are four scenarios for short-range spin glasses which have (so far) been found to be consistent mathematically~\cite{NS03a,NS22,NRS23b} and which agree with common findings of almost all numerical simulations to date --- in particular,  the existence of a pair of $\delta$-functions in the spin overlap distribution at $\pm q_{EA}$, where $q_{EA}$ is the Edwards-Anderson order parameter~\cite{EA75}. Of these, three (droplet-scaling~\cite{Mac84,BM85,BM87,FH86,FH88b}, TNT~\cite{MP00,PY00}, and chaotic pairs~\cite{NS96c,NS97,NSBerlin,NS03b}) propose that spin glass thermodynamic states (one in the cases of droplet-scaling and TNT, many in the case of chaotic pairs) are each a trivial mixture of a single spin-reversed pure state pair, while the fourth (replica symmetry breaking, or RSB~\cite{Parisi79,Parisi83,MPSTV84a,MPSTV84b,MPV87}) requires that thermodynamic states comprise nontrivial mixtures of a countably infinite set of incongruent (to be defined below) pure state pairs. 

In this note we construct a new type of metastate, which we call the {\it restricted metastate\/}, that groups together pure states having a common overlap with a reference pure state. The restricted metastate allows us to study fluctuations of free energy differences between incongruent pure states. These fluctuations are fundamental to understanding the nature of the spin glass phase in finite dimensions; indeed, frustration was originally defined in terms of these fluctuations~\cite{Anderson78}, and they form the basis for defining and computing the spin glass stiffness. Perhaps most importantly, determining their behavior provides a means to distinguish among the possible scenarios for the spin glass phase described above.

Detailed proofs of the claims made in this note will be published elsewhere~\cite{NSinprep}; here we present the main ideas and provide an informal sketch of the main constructions and arguments.


We consider the Edwards-Anderson (EA) nearest-neighbor Ising spin glass model~\cite{EA75} in zero magnetic field on the $d$-dimensional cubic lattice $\mathbb{Z}^d$ with Hamiltonian
\begin{equation}
\label{eq:EA}
{\cal H}_J=-\sum_{(x,y)} J_{xy} \sigma_x\sigma_y\, , 
\end{equation}
where $\sigma_x=\pm 1$ is the Ising spin at site $x$ and $(x,y)$ denotes an edge in the (nearest-neighbor) edge set $\mathbb{E}^d$. The couplings $J_{xy}$ are independent, identically distributed random variables chosen from a distribution $\nu(dJ_{xy})$, with random variable $J_{xy}$  assigned to the edge $(x,y)$; while the arguments presented here apply to a general class of distributions with mild restrictions~\cite{NSinprep}, for specificity in this paper we will take $\nu$ to be Gaussian with mean zero and variance one. We denote by $J$ a particular realization of the couplings. Because of the spin-flip symmetry of the Hamiltonian, any mixed thermodynamic state generated by a spin-symmetric boundary condition (e.g., free or periodic) will be supported on spin-reversed pure state pairs (hereafter denoted simply as pure state pairs).

Using this Hamiltonian we consider a metastate $\kappa_J$ at some fixed temperature~$T$ using either the Aizenman-Wehr (AW)~\cite{AW90} or Newman-Stein (NS)~\cite{NS96c} approach. A metastate is a probability measure on infinite-volume Gibbs states, which in turn are probability measures on infinite-volume spin configurations (see~\cite{NRS23b} for a review). In particular, suppose one examines an infinite sequence of volumes $\Lambda_L=[-L,L]^d\subset\mathbb{Z}^d$ all centered at the origin and each with a specified boundary condition.  Depending on the Hamiltonian, temperature, and boundary conditions chosen, this sequence of finite-volume Gibbs states might converge to a single (pure or mixed) infinite-volume Gibbs state (i.e., a thermodynamic state), or else it may not converge but have two or more subsequences converging to different Gibbs states. The metastate is a probability measure that describes the distribution of these distinct thermodynamic states, or equivalently, it describes the collection of all correlation functions within a large arbitrary volume. Informally, it provides information on the fraction of volumes converging to one or another of the limiting thermodynamic states.






More formally, a metastate is a probability measure on infinite-volume Gibbs states, depending on $J$ and $\beta$, satisfying the properties of {\it coupling-covariance\/} and {\it translation-covariance\/}.  The latter simply requires that a uniform lattice shift does not affect the metastate properties, and can be expressed as the following requirement: for any lattice translation $\tau$ of $\mathbb{Z}^d$ and a subset $A$ of probability measures on the space of spin configurations $\{-1,+1\}^{\mathbb{Z}^d}$,
\begin{equation}
\label{eq:2}
\kappa_{\tau J}(A)=\kappa_{J}(\tau^{-1}A).
\end{equation}
Translation-covariance is guaranteed when one constructs a metastate using periodic boundary conditions to generate the finite-volume Gibbs states; in the infinite-volume limit, the Gibbs states (and therefore the metastate) will inherit the torus-translation covariance of the finite-volume Gibbs states. Hereafter we will always consider periodic boundary condition~(PBC) metastates as our starting point. 


Coupling covariance refers to transformations on states under a finite change of couplings, i.e., the values of a finite number of couplings are changed by finite amounts.  Changing a finite set of couplings of course will change the thermodynamic states, in that the correlation functions determined by the state will also change. However, it can be shown that under a finite change of couplings, a pure state transforms to a pure state~\cite{AW90,NS96c}, and therefore a convex mixture of multiple pure states (i.e., a mixed Gibbs state) will remain a convex mixture of the transformed pure states, generally with modified weights.  Coupling covariance can then be expressed as follows:  for~$B$ a finite subset of $\mathbb{Z}^d$, $J_B$ the set of couplings assigned to the edges in $B$, $f(\sigma)$ a function of a finite set of spins, and $\Gamma$ a Gibbs state, we define the operation ${\cal L}_{J_B} : \Gamma \mapsto {\cal L}_{J_B}\Gamma$ by its effect on the expectation $\langle\cdots\rangle_\Gamma$ 
in $\Gamma$:
\begin{equation}
\label{eq: gamma L}
 \left\langle f(\sigma)\right\rangle_{{\cal L}_{J_B}\Gamma}= \frac{\left\langle f(\sigma) \exp\Bigl(-\beta H_{J_B}(\sigma)\Bigr)\right\rangle_\Gamma}
{\left\langle\exp\Bigl(-\beta H_{J_B}(\sigma)\Bigr)\right\rangle_\Gamma}\, ,
\end{equation}
which describes the effect of modifying the couplings within $B$. We require that the metastate 
be covariant under local modifications of the couplings, i.e., for any subset $A$ defined as in~(\ref{eq:2}),
\begin{equation}
\label{eq:3}
\kappa_{J+J_B}(A)= \kappa_{J}({\cal L}_{J_B}^{-1}A)\, ,
\end{equation}
where ${\cal L}_{J_B}^{-1} A$ equals the set of $\Gamma$'s such that ${\cal L}_{J_B}\Gamma\in A$.
In other words, the set of Gibbs states on which the metastate is supported does not change, aside from the usual changes in correlation functions within the individual states.
In particular, no Gibbs states either flow into or out of the metastate under a finite change of couplings.

Proceeding, let $E_L = \mathbb{E}(\Lambda_L)$, the edge set within $\Lambda_L$, and define the edge (or bond) overlap between two states $\alpha$ and $\beta$ as
\begin{equation}
\label{eq:overlap}
q^{(e)}_{\alpha\beta}=\lim_{L\to\infty}\frac{1}{d\vert\Lambda_L\vert}\sum_{\langle xy\rangle\in E_L}\langle\sigma_x\sigma_y\rangle_\alpha\langle\sigma_x\sigma_y\rangle_\beta\, .
\end{equation}

The following result will be important in what follows: Given a coupling realization~$J$, at any fixed positive temperature the bond overlap $q^{(e)}_{\alpha\beta}$ is invariant under a finite change in couplings.
The proof is long and will appear elsewhere~\cite{NSinprep}, and also can be found online~\cite{noteNS}.

We next introduce the notion of incongruence~\cite{HF87,FH87}, using the following definition from~\cite{ANSW14}.  Two pure states $\alpha$ and $\beta$ are defined to be incongruent if for some $\epsilon>0$ there is a subset of edges $(x_0,y_0)$ with strictly positive density such that $\vert\langle\sigma_{x_0}\sigma_{y_0}\rangle_\alpha-\langle\sigma_{x_0}\sigma_{y_0}\rangle_\beta\vert > \epsilon$.
This is equivalent to the condition $q^{(e)}_{\alpha\beta}<q^{(e)}_{\alpha\alpha}$, where the self-overlap~$q^{(e)}_{\alpha\alpha}$ is the same for all pure states within a single mixed Gibbs state~\cite{NRS23a}.

We note that if a PBC~metastate is supported on nontrivial mixed Gibbs states (i.e., Gibbs states which are mixtures of more than one pure state pair), then~(1) all non-spin-flip-related pure states in the metastate are mutually incongruent~\cite{NS01c}, and (2) the metastate barycenter (also called the metastate averaged state~\cite{Read14}) is  a single mixed Gibbs state whose decomposition into pure states has no atoms~\cite{NS06b,Read14}.

We now construct a new type of metastate which will be referred to as a {\it restricted metastate\/}.  Begin by choosing a pure state $\omega$ randomly from~$\kappa_J$ and an interval $(p-\delta,p+\delta)$ with $p\in(-1,1)$ and $\delta>0$.  We use two different constructions which may give rise to different restricted metastates, but which will both satisfy the three properties listed earlier.

{\it Construction~1\/.}  For every Gibbs state $\Gamma$ in $\kappa_J$ consider the edge overlap $q^{(e)}_{\alpha\omega}$ for every pure state $\alpha$ in the decomposition of $\Gamma$. If $q^{(e)}_{\alpha\omega}\in(p-\delta,p+\delta)$ we retain $\alpha$, otherwise $\alpha$ is discarded. (If no pure state in $\Gamma$ satisfies this condition, $\Gamma$ itself is discarded.) The weights of the retained $\alpha$'s are rescaled so that the new weights add up to one with their relative probabilities the same as before. Finally we renormalize the overall mass to compensate for the discarded~$\Gamma$'s. 

It could be (see, e.g.~\cite{Read14}) that the overlap distribution of the barycenter of $\kappa_J$ is a single $\delta$-function, in which case Construction~1 may (depending on $p,\delta$) discard a set of $\Gamma$'s with measure one in the metastate.  To avoid this situation, one would then use Construction~2.

{\it Construction 2.\/}  For each $\omega$ one retains only the $\Gamma$ from which $\omega$ was chosen~\cite{omeganote}, and then follows the procedure of Construction~1 for each of the pure states in $\Gamma$.

The discussion so far outlines a procedure that uses a fixed $\omega$ chosen from the PBC metastate. In order to construct a new metastate, $\omega$ itself is treated as a random variable.
The resulting object is a~$(p,\delta)$-restricted measure ${\kappa}^{p,\delta}_{J,\omega}$ on Gibbs states (the notation is chosen to separate $p$ and $\delta$, which are fixed parameters, from $J$ and $\omega$, which are random quantities). If the PBC~metastate is supported on an uncountable set of states, then in both constructions the restricted metastate is also, albeit in a novel way as $\omega$ varies.

At any positive temperature, ${\kappa}^{p,\delta}_{J,\omega}$ as constructed above satisfies the three conditions for a translation-covariant metastate (but now depending on $J$, $\omega$, and $\beta$).  To see this, note that by the method of construction, ${\kappa}^{p,\delta}_{J,\omega}$ is supported solely on mixtures of pure states appearing in the support of the PBC metastate, and the renormalization of its mass guarantees that the probabilities of all states add up to one.  As noted earlier, overlaps are invariant with respect to finite changes in couplings, which ensures that coupling covariance, which now takes the form
\begin{equation}
\label{eq:cc}
{\kappa}^{p,\delta}_{J+J_B,{\cal L}_{J_B}\omega}(A)= {\kappa}^{p,\delta}_{J,\omega}({\cal L}_{J_B}^{-1}A)\, ,
\end{equation}
is satisfied.  

Translation-covariance follows because $\omega$ is treated as a random variable rather than as a fixed state; i.e., an event, which in this setting is a function on the spins, is evaluated in terms of its $(J,\omega)$-probability.  Eq.~(\ref{eq:3}) is then replaced by
\begin{equation}
\label{eq:modcov}
{\kappa}^{p,\delta}_{\tau J,\tau\omega}(A)={\kappa}^{p,\delta}_{J,\omega}(\tau^{-1}A)
\end{equation}
which, given the translation covariance of Gibbs states in $\kappa_J$, is clearly satisfied. 
%

In the above constructions it was assumed that there exist multiple edge overlap values for incongruent pure states chosen from a mixed state~$\Gamma$ in~$\kappa_J$; this is an important feature of replica symmetry breaking~\cite{Parisi79,Parisi83,MPSTV84a,MPSTV84b,MPV87} in which spin and edge overlaps from such a~$\Gamma$ span a range of values.  In contrast, the PBC metastate~$\kappa_J$ for droplet-scaling and TNT comprises a single trivial mixed Gibbs state supported on a pair of spin-reversed pure states with equal weight. In the chaotic pairs picture $\kappa_J$ is supported on multiple mixed Gibbs states, but each of these is again a trivial mixture of a single spin-reversed pure state pair. In droplet-scaling and TNT (and chaotic pairs when Construction 2 is used), the restricted metastate will either be empty (if $q^{(e)}_{EA}\notin[p-\delta,p+\delta]$) or else it will simply map $\kappa_J$ back to itself.  

Thus the restricted metastate construction is primarily useful in scenarios, such as RSB, where~$\kappa_J$ is supported on nontrivial mixed Gibbs states.

From here on we write $(J,\omega)$ for a random pair consisting of a coupling realization and a (random) choice of $\omega$ for that individual~$J$.  As before, we assume that the PBC metastate~$\kappa_J$ is supported on multiple incongruent pure state pairs with a nontrivial overlap distribution.  Let $\overline{f(\sigma_x\sigma_y)}=\lim_{L\to\infty}\frac{1}{dL^d}\sum_{\langle xy\rangle\in E_L} f(\sigma_x\sigma_y)$ denote the spatial average of a measurable function~$f$ of edge variables.  Clearly, given a PBC metastate $\kappa_J$, for each~$(J,\omega)$ pair $\overline{\langle\sigma_x\sigma_y\rangle_\Gamma\langle\sigma_x\sigma_y\rangle_{\omega}}=p+O(\delta)$ for any Gibbs state~$\Gamma$ in the support of the restricted metastate~${\kappa}^{p,\delta}_{J,\omega}$. (A full proof will be given in~\cite{NSinprep}, but the result follows straightforwardly by the method of construction of the restricted metastate.)

In what follows it is convenient to use the notation ${\tilde\kappa}^{p,\delta}_{J,{\hat\omega}}$ when the $\omega$ in ${\kappa}^{p,\delta}_{J,\omega}$ is no longer random but equals a particular $\hat\omega$.  Then consider ${\tilde\kappa}^{p_1,\delta}_{J,\hat\omega}$ and ${\tilde\kappa}^{p_2,\delta}_{J,\hat\omega}$ chosen from $\kappa_J$ with the same $\hat\omega$, and with $0\le p_1<p_2<1$ and $0<\delta< {\rm min}(p_1,p_2-p_1, 1-p_2)$.  We then have 
 \begin{eqnarray}
 \label{eq:r1}
{\tilde\kappa}^{p_1,\delta}_{J,\hat\omega}\Bigl(\overline{\langle\sigma_x\sigma_y\rangle_\Gamma\langle\sigma_x\sigma_y\rangle_{\hat\omega}}\Bigr)=p_1 + O(\delta)\nonumber\\
\ne{\tilde\kappa}^{p_2,\delta}_{J,\hat\omega}\Bigl(\overline{\langle\sigma_x\sigma_y\rangle_\Gamma\langle\sigma_x\sigma_y\rangle_{\hat\omega}}\Bigr)=p_2 + O(\delta)\, .
\end{eqnarray}

The inequality~(\ref{eq:r1}) can hold only if ${\tilde\kappa}^{p_1,\delta}_{J,\hat\omega}\Bigl(\langle\sigma_x\sigma_y\rangle_\Gamma\langle\sigma_x\sigma_y\rangle_{\hat\omega}\Bigr)\ne{\tilde\kappa}^{p_2,\delta}_{J,\hat\omega}\Bigl(\langle\sigma_x\sigma_y\rangle_\Gamma\langle\sigma_x\sigma_y\rangle_{\hat\omega}\Bigr)$ for a positive density of edges. Because the choice of $\hat\omega$ is the same for both metastates, it must then also be true that ${\tilde\kappa}^{p_1,\delta}_{J,\hat\omega}(\langle\sigma_x\sigma_y\rangle_\Gamma)\ne{\tilde\kappa}^{p_2,\delta}_{J,\hat\omega}(\langle\sigma_x\sigma_y\rangle_\Gamma)$ for a positive density of edges. Because this is so for each~instance of $(J,\hat\omega)$, it follows that for any edge~$(x,y)$
\begin{equation}
\label{eq:incong2}
(\nu\times\kappa_J)\Bigl\{(J,\omega):{\kappa}^{p_1,\delta}_{J,\omega}\big(\langle\sigma_x\sigma_y\rangle_\Gamma\big)\neq {\kappa}^{p_2,\delta}_{J,\omega}\big(\langle\sigma_x\sigma_y\rangle_\Gamma\big)\Bigr\}>0\, ,
\end{equation}
where $\nu\times\kappa_J$ denotes $\nu(dJ)\kappa_J(d\omega)$.

The preceding discussion demonstrates that the restricted metastate satisfies the conditions for Theorem~4.2 of~\cite{ANSW14} to apply. In the present context this can be expressed as: 

\medskip

{\bf Theorem} (modified from~\cite{ANSW14}):  Consider two infinite-volume (pure or mixed) Gibbs states $\Gamma$ and $\Gamma'$ chosen from distinct restricted metastates satisfying~(\ref{eq:incong2}), and let  $F_L(\Gamma,\Gamma')$ denote their free energy difference restricted within a volume~$\Lambda_L=[-L,L]^d\subset\mathbb{Z}^d$ (for a formal definition, see Eq.~(3) in~\cite{ANSW14}) . Then 
there is a constant $c>0$ such that the variance of $F_L(\Gamma,\Gamma')$ under the probability measure $M:=\nu(dJ)\kappa_J(d\omega){\kappa}^{p_1,\delta}_{J,\omega}(d\Gamma)\times{\kappa}^{p_2,\delta}_{J,\omega}(d\Gamma')$ satisfies
\begin{equation}
\label{eq:flucs}
{\rm Var}_M\Big(F_L(\Gamma,\Gamma')\Big)\ge c\vert\Lambda_L\vert\, .
\end{equation}

This result demonstrates that a large variance, i.e., one scaling with the volume, occurs in the free energy difference of two incongruent pure states within the same mixed~$\Gamma$. To see this, suppose first that the spin glass phase is described by one-step RSB, so that in each mixed Gibbs state~$\Gamma$ there are only two edge overlap values: the self-overlap $q^{(e)}_{EA}$ and an overlap $q_0(<q^{(e)}_{EA})$ between two incongruent pure states in the decomposition of~$\Gamma$. Then using construction~2 one constructs two incongruent restricted metastates by choosing $p_1=q^{(e)}_{EA}$ and $p_2=q_0$.  Next suppose that in a given~$\Gamma$ there were two distinct (non-self) overlap values, $p_1$ and $p_2$. Then one can always find three incongruent pure states $\alpha$, $\beta$, and $\omega$ in that~$\Gamma$ for which $q^{(e)}_{\alpha\omega}=p_1$ and $q^{(e)}_{\beta\omega}=p_2$, and similarly for $k$-step RSB with arbitrary $k$. In all of these cases any two (incongruent) pure states $\alpha$ and $\beta$ can be chosen to belong to different restricted metastates as in~(\ref{eq:incong2}), and so the lower bound~(\ref{eq:flucs}) applies.

There is also an upper bound following from a known result on the free energy itself~\cite{WA90}, which when applied to the situation considered here can be stated as
\begin{equation}
\label{eq:flucsupper}
{\rm Var}_M\Big(F_L(\Gamma,\Gamma')\Big)\le d\vert\Lambda_L\vert\, ,
\end{equation}
where $d>0$ is again positive and independent of the volume.

We have thus shown the following:

{\bf Main Result.} Suppose that in the EA Ising model there is a phase consisting of nontrivial mixed states. Then for any two pure states $\alpha$ and $\alpha'$ within a single mixed Gibbs state~$\Gamma$ in the PBC metastate~$\kappa_J$, the fluctuations of their difference in free energy ${\rm Var}_M\Big(F_L(\alpha,\alpha')\Big)$ scales with the volume.

We now turn to a heuristic discussion of the implications of these results (which can be extended to coupling-independent boundary conditions besides periodic, though we shall not pursue that here).  We emphasize that our result applies to {\it finite-volume restrictions of infinite-volume Gibbs states\/}. If one focuses on finite systems, as opposed to finite-volume restrictions of infinite systems, an analog of the theorem provides information on free energy difference fluctuations inside a window far from the boundaries, but not over the entire volume (or indeed most of it).  


A simple scenario consistent with the theorem is that the spin glass phase does not contain nontrivial mixed states, as in droplet-scaling, TNT, or chaotic pairs.  In these scenarios any mixed state consists of only a single pair of spin-reversed pure states, each with equal free energy in any restricted volume.


There are also lines of reasoning that allow the theorem to coexist with RSB.  As noted earlier, in RSB a single infinite-volume mixed Gibbs state is supported on infinitely many pure states. In any convergent subsequence of finite-volume Gibbs states, none of the (finite-volume approximations to the) pure states can have free energy differences that scale with the volume, or else they wouldn't appear in the same mixed Gibbs state. But there could still be a {\it sub\/}sequence of volumes in which the free energy difference between pure states $\alpha$ and $\alpha'$ are all $O(1)$, e.g., when $F_L(\alpha,\alpha')$ changes sign. (Of course one then require the existence of subsubsequences of volumes in which {\it all\/} pure states within a single mixed Gibbs state have free energy differences of $O(1)$.)  So even if two pure states $\alpha$ and $\alpha'$ appear in the same mixed state~$\Gamma$, they might nonetheless have large free energy difference fluctuations inside {\it most\/} restricted volumes~\cite{noteRead}, but not all. Our results do not rule out this possibility. But this would require that the (infinite-volume) interface $\alpha\triangle\alpha'$ have a free energy that scales as $L^{d/2}$ (while changing sign infinitely often).

In this note we have discussed only pure states within a single mixed state~$\Gamma$.  One can also extend these results to the free energy differences between pure states from different mixed states, between entire mixed states, and between energy difference fluctuations among ground states. These results will be reported elsewhere~\cite{NSinprep}.


\smallskip


{\it Acknowledgments.\/} We thank Louis-Pierre Arguin, Aernout van Enter, and Jon Machta for useful discussions and suggestions.  We are especially indebted to Nick Read for an illuminating and fruitful correspondence that improved earlier versions of the paper.

\bibliography{refs}

\begin{thebibliography}{34}
\bibitem [{ome()}]{omeganote}%
  \BibitemOpen
  \href@noop {} {}\bibinfo {note} {This procedure assumes that in the case
  under consideration the chosen $\omega$ belongs to a single $\Gamma$.
  Otherwise choose a $(\Gamma,\omega)$ pair instead of just
  $\omega$.}\BibitemShut {Stop}%
\end{thebibliography}
\end{document}